\documentstyle[11pt,aaspp4]{article}

\input{psfig.sty}

\begin{document}

\title{The 2001 April Burst Activation of SGR~1900$+$14: Pulse Properties and
Torque}

\author{
P.M.~Woods\altaffilmark{1,2},
C.~Kouveliotou\altaffilmark{2,3},
E.~{G\"o\u{g}\"u\c{s}}\altaffilmark{1,2},
M.H.~Finger\altaffilmark{1,2},
M.~Feroci\altaffilmark{4},
S.~Mereghetti\altaffilmark{5},
J.H.~Swank\altaffilmark{6}, 
K.~Hurley\altaffilmark{7},
J.~Heise\altaffilmark{8},
D.~Smith\altaffilmark{9},
F.~Frontera\altaffilmark{10,11},
C.~Guidorzi\altaffilmark{11}, and
C.~Thompson\altaffilmark{12}
}

\altaffiltext{1}{Universities Space Research Association; 
peter.woods@nsstc.nasa.gov, ersin.gogus@nsstc.nasa.gov,
mark.finger@nsstc.nasa.gov}
\altaffiltext{2}{National Space Science and Technology Center, 320 Sparkman Dr. 
Huntsville, AL 35805}
\altaffiltext{3}{NASA Marshall Space Flight Center, Huntsville, AL 35812;
chryssa.kouveliotou@nsstc.nasa.gov}
\altaffiltext{4}{Istituto di Astrofisica Spaziale e Fisica Cosmica - CNR, Rome,
Italy; feroci@rm.iasf.cnr.it}
\altaffiltext{5}{Istituto di Astrofisica Spaziale e Fisica Cosmica - CNR,
Sezione di Milano, Italy; sandro@mi.iasf.cnr.it}
\altaffiltext{6}{NASA Goddard Space Flight Center, Greenbelt, MD 20771; 
swank@pcasun1.gsfc.nasa.gov}
\altaffiltext{7}{University of California at Berkeley, Space Sciences
Laboratory, Berkeley, CA 94720-7450; khurley@ssl.berkeley.edu}
\altaffiltext{8}{SRON National Institute for Space Research, Sorbonnelaan 2, 
3584 CA Utrecht, The Netherlands; jheise@purple.sron.nl}
\altaffiltext{9}{Department of Physics, University of Michigan at Ann Arbor, 
501 East University Avenue, Ann Arbor, MI 48109; 
dasmith@rotse2.physics.lsa.umich.edu}
\altaffiltext{10}{Istituto di Tecnologie e Studio delle Radiazioni 
Extraterrestri (CNR), Via Gobetti 101, 40129 Bologna, Italy; 
filippo@tesre.bo.cnr.it}
\altaffiltext{11}{Dipartimento Fisica, Universita di Ferrara, Via Paradiso 12, 
44100, Ferrara, Italy; guidorzi@fe.infn.it}
\altaffiltext{12}{Canadian Institute for Theoretical Astrophysics, 60 St.\
George Street, Toronto, ON M5S 3H8, Canada; thompson@cita.utoronto.ca}

\begin{abstract}

We report on observations of SGR~1900$+$14 made with the {\it Rossi X-ray
Timing Explorer (RXTE)} and {\it BeppoSAX} during the April 2001 burst
activation of the source.  Using these data, we measure the spindown torque on
the star and confirm earlier findings that the torque and burst activity are
not directly correlated.  We compare the X-ray pulse profile to the gamma-ray
profile during the April 18 intermediate flare and show that ($i$) their shapes
are similar and ($ii$) the gamma-ray profile aligns closely in phase with the
X-ray pulsations.  The good phase alignment of the gamma-ray and X-ray profiles
suggests that there was no rapid spindown following this flare of the magnitude
inferred for the August 27 giant flare.  We discuss how these observations
further constrain magnetic field reconfiguration models for the large flares of
SGRs.

\end{abstract}


\newpage

\section{Introduction}

Soft Gamma Repeaters (SGRs) are a small class (4 known) of intriguing
high-energy transient that emit anywhere from a handful to several hundred
brief, intense bursts of soft gamma-rays when active (see Kouveliotou 2003 for
a recent review).  In quiescence, SGRs have been found to exhibit persistent
X-ray luminosities of $\sim$10$^{35}$ ergs s$^{-1}$.  Three SGRs emit coherent
pulsations between 5 and 8 s (Kouveliotou et al.\ 1998; Hurley et al.\ 1999a;
Kulkarni et al.\ 2003), and all are spinning down rapidly with time
(Kouveliotou et al.\ 1998, 1999, Kulkarni et al.\ 2003).  The spindown observed
in two of these SGRs (1806$-$20 and 1900$+$14) is not constant.  Furthermore,
most of the spindown variations do not directly correlate with the bursting
activity of the source (Woods et al.\ 2002).

The rapid spindown in SGRs has been interpreted (e.g.\ Kouveliotou et al.\
1998) as magnetic braking of a strongly magnetized neutron star with $B_{\rm
dip}$ $\sim 10^{14} - 10^{15}$ G, or magnetar (Thompson \& Duncan 1995, 1996). 
The magnetar model postulates that the short duration SGR bursts are triggered
by starquakes induced by magnetic stresses in the neutron star crust (Thompson
\& Duncan 1995) or perhaps magnetic reconnection events in the stellar
magnetosphere (Lyutikov 2002).  Persistent magnetospheric currents, driven by
twists in the evolving magnetic field, and field decay in the stellar interior
contribute to the quiescent flux from SGRs (Thompson \& Duncan 1996; Thompson,
Lyutikov \& Kulkarni 2002).

Burst activity in SGRs occurs sporadically in time and the separation between
successive events can vary from seconds to years.  Epochs when SGRs are
emitting several bursts or burst active phases vary in both intensity and
duration.  For example, SGR~1900$+$14 was discovered in 1979 when it was
observed to burst only 3 times in 3 days (Mazets \& Golenetskii 1981).  In
1992, the source became active again for a few days and emitted a handful of
events (Kouveliotou et al.\ 1993); it entered an unprecedented level of
activity in 1998 (Hurley et al.\ 1999b) never before observed for any SGR. 
During the course of 9 months, more than 1000 bursts were recorded from this
source with various instruments ({G\"o\u{g}\"u\c{s}} et al.\ 1999).

The pinnacle of the 1998 activity of SGR~1900$+$14 was realized on August 27
when a giant flare was recorded from this source (Hurley et al.\ 1999c; Mazets
et al.\ 1999; Feroci et al.\ 1999, 2000).  The event reached a peak luminosity
of $\sim$4 $\times$ 10$^{44}$ ergs s$^{-1}$ and persisted for over $\sim$6
minutes releasing a total of $\sim$10$^{44}$ ergs in gamma-rays $>$15 keV.  The
flare started with a spectrally hard, non-thermal initial spike that was
followed by a softer tail which decayed in a quasi-exponential manner. 
Superimposed on the decaying tail were coherent 5.16 s pulsations whose shape
evolved with time.  Eighteen minutes after the termination of the gamma-ray
emission, a bright X-ray tail from SGR~1900$+$14 was detected that decayed in
time as a power-law (Woods et al.\ 2001).  The change in pulse profile observed
during the August 27 flare at gamma-ray energies was also evident in the
persistent X-ray emission pulse profile seen before and after the flare (Woods
et al.\ 2001; {G\"o\u{g}\"u\c{s}} et al.\ 2002).  In addition, a transient
radio outburst was seen for the first and only time from an SGR source, due to
a sudden outflow of material associated with this flare (Frail, Kulkarni \&
Bloom 1999).  Finally, there is strong circumstantial evidence which indicates
that there was a brief, albeit substantial spin-down event ($\Delta$P =
5.72(14) $\times$ 10$^{-4}$ s [Woods et al.\ 1999b]) within the hours following
the flare (Palmer et al.\ 2001), due perhaps to this large outflow of material
(Thompson et al.\ 2000).

On 2001 April 18, a burst was detected from SGR~1900$+$14 with a high energy
($\sim$10$^{43}$ ergs) and long duration $\sim$40 s (Guidorzi et al.\ 2001). 
Unlike the August 27 flare, there was no intense, non-thermal emission episode
at the onset of this event (Guidorzi et al.\ in preparation).  The energy
released by this flare was less than the giant flare of August 27, yet much
larger than a typical SGR event and was consequently dubbed an ``intermediate
flare'' (Kouveliotou et al.\ 2001).  Several, more common SGR bursts were
detected during the following weeks.  We triggered a sequence of ToO
observations of the source with the {\it Rossi X-ray Timing Explorer} ({\it
RXTE}) and the {\it Satellite per Astronomia X} ({\it BeppoSAX}) over the two
weeks following the April 18 event.  Here, we present pulse timing and profile
results from these observations.  The flux history during this epoch is
discussed in the accompanying paper (Feroci et al.\ 2003).

\section{Observations}

The ToO observations of SGR~1900$+$14 with the {\it RXTE} Proportional Counter
Array (PCA) began $\sim$34 hours after the intermediate flare and continued for
the next two weeks.  A total of 128 ks of data were collected during 13
pointings.  A single monitoring observation (10 ks exposure time) of
SGR~1900$+$14 was serendipitously performed on 2001 April 14, just 4 days prior
to the flare.

We find no burst activity during the pre-flare observation on 2001 April 14. 
We performed a search through the data following the flare and found a total of
32 bursts that exceeded 5.5$\sigma$ above background (using a running mean) on
the 0.125 s time scale.  The last burst from SGR~1900$+$14 detected in the PCA
data was recorded on 2001 May 1 at 04:00:22 UT.  On 2001 April 28, we noted a
sudden increase in the intensity of SGR~1900$+$14 between consecutive {\it
RXTE} orbits.  We determined that this flux increase was associated with
another high fluence burst from SGR~1900$+$14 recorded by instruments aboard
{\it Ulysses} and {\it Konus} while the source was Earth-occulted for {\it
RXTE}.  A detailed analysis of this event is presented elsewhere (Lenters et
al.\ 2003).

Two observations of SGR~1900$+$14 were performed with the {\it BeppoSAX} Narrow
Field Instruments (NFI) during the April 2001 activation of the source.  The
first commenced $\sim$8 hours after the intermediate flare and the second
observation took place $\sim$12 days later.  The source exposure times for the
two Medium Energy Concentrator Spectrometer (MECS) units were 34.8 and 57.3 ks,
respectively.  A more detailed account of the NFI observations are presented in
Feroci et al.\ (2003).

\section{Pulse Timing Analysis}

The PCA data from each observation were first filtered to remove the 32
detected bursts by eliminating all data prior to and following each burst by 1
s.  The data were then energy selected (2$-$5 keV) to maximize the
signal-to-noise ratio and allow for a comparison to the {\it BeppoSAX} NFI
data.  The data were binned to 0.125 s time resolution and transformed to the
Solar system barycenter using the {\bf ftool} {\it faxbary}.  Next, we
generated Lomb-Scargle power spectra (Lomb 1976) to determine coarse
frequencies for each set of observations.  

The {\it BeppoSAX} MECS data were filtered for bursts by first binning the
source region event times into a histogram having 0.5 s time resolution and
calculating the normalized Poisson probabilities for each bin.  Bins having
probabilities less than 1 $\times$ 10$^{-3}$ ($\sim$7 counts) were identified
as bursts and events recorded during those times ($\pm$1 s) were removed from
the subsequent analysis.  As with the PCA data, we energy selected our event
list to only include photons with energies 2$-$5 keV.  The event times were
transformed to the solar-system barycenter using the SAXDAS tool {\it
baryconv}.

All data following the flare were folded on a single frequency determined by
the highest peak in the Lomb-Scargle power spectrum and a phase was calculated
for each set relative to a template profile (generated from a subset of PCA
observations when the source was brightest).  We found significant curvature in
the phase residuals, indicative of spin-down.  The phases were fit to a second
order polynomial and a new template was generated from the full data set.  This
procedure was iterated and the final second-order polynomial fit yielded a
$\chi^2=$ 20.8 for 26 degrees of freedom.  Using the combined {\it RXTE} and
{\it BeppoSAX} data sets, we measure a frequency and frequency derivative of
0.193317095(15) Hz and  $-$6.56(9) $\times$ 10$^{-12}$ Hz s$^{-1}$,
respectively (epoch = 52023.0 MJD TDB).  The combined data set spans the time
range MJD 52017.68$-$52034.01 (Figure 1).  The spindown rate measured here is
more rapid than the spindown observed following the giant flare of 1998 August
27 (Woods et al.\ 1999b) by a factor $\sim$2.9.


\begin{figure}[!htb]
\centerline{
\psfig{file=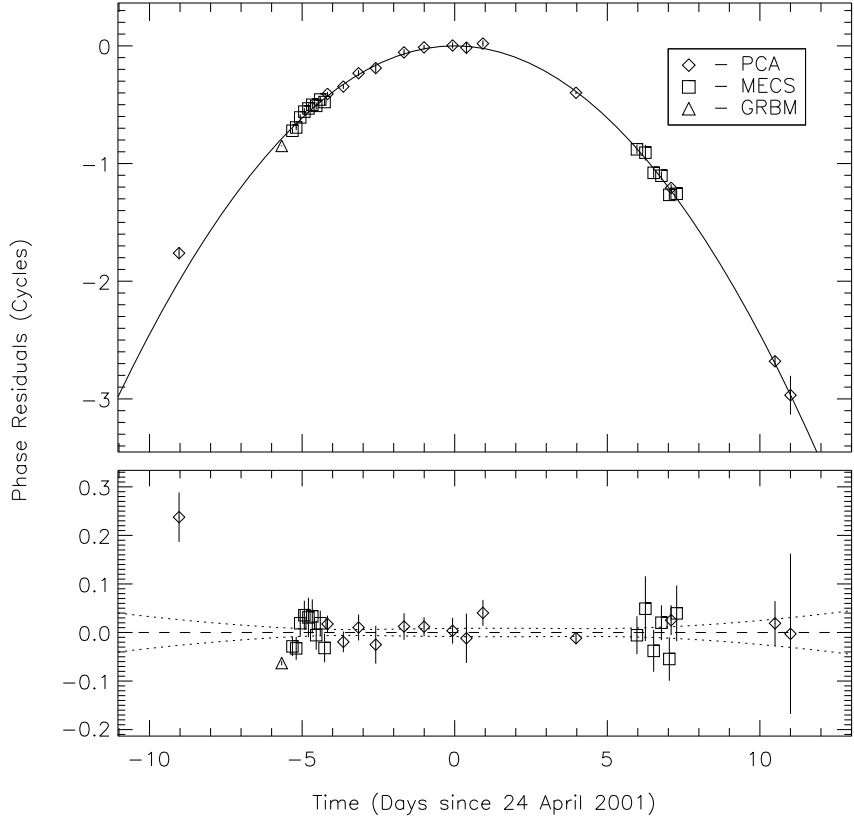,height=4.8in}}
\vspace{-0.05in}

\caption{{\it Top} -- Pulse phases (PCA, MECS, and GRBM) during the April
activation of SGR~1900$+$14 minus a linear trend. {\it Bottom} -- The same
phases minus a 2$^{\rm nd}$ order polynomial fit to the PCA and MECS phases
after April 18.  The dotted lines represent the $\pm$1$\sigma$ errors to this
fit.  Note the outlier at $\sim-$9 days preceding the April 18 flare
(triangle).}

\vspace{11pt}
\end{figure}

We next extrapolated our fit to the time of the PCA observation that preceded
the April 18 flare by four days.  Using the full covariance matrix from our fit
to the post-flare phases, the model uncertainty in the predicted phase at the
time of the pre-flare observation is $\pm$0.027 cycles.  We find that the
relative phase measured for this observation is discrepant from the model
ephemeris by $\sim$0.24 cycles or 4.1$\sigma$.  The frequency during this brief
observation, however, is poorly determined so we cannot rule out the existence
of cycle slips between April 14 and April 18 which would only increase the
phase discrepancy (e.g.\ the phase difference could be -1.76, -0.76, 0.24,
1.24, cycles, etc.).  The disagreement of our phase measurement on April 14
suggests that the source suffered atiming anomaly, however, the
absence of a precise frequency measurement on April 14 precludes us from
determining the manner in which the spin evolution deviated.  Therefore, it is
not possible to place a meaningful quantitative limit on any hypothetical
glitch coincident with or following the April 18 flare.

Palmer (2001) noted that the gamma-ray pulsations (15$-$150 keV) during the
1998 August 27 flare were $\sim$150$^{\circ}$ out of phase with the post-flare
X-ray pulse ephemeris (2$-$10 keV) measured one day after the flare.  He noted
that this mis-alignment could be either due to a strong energy dependence of
the pulse profile or the consequence of rapid spin-down during the minutes or
hours following the flare as was deduced earlier from the spin history of the
source (Woods et al.\ 1999).  Using the GRBM data from {\it BeppoSAX}, we
performed a similar analysis for the April 18 flare.  The GRBM bin times were
converted to the solar-system barycenter to within the absolute time accuracy
of the GRBM clock ($\sim$10 ms or 0.002 cycles).  Using the spin ephemeris
determined above, we calculated the phase of each GRBM time bin.  The model
uncertainty in the phase at the time of the flare is 0.001 cycles.  The
detrended flare data are plotted along with the folded (2$-$5 keV) PCA X-ray
profile (Figure 2).  Unlike the August 27 flare, we find good agreement in the
occurrence in phase of the peak in the pulse profile between the X-ray and
gamma-ray bands.


\begin{figure}[!htb]
\centerline{
\psfig{file=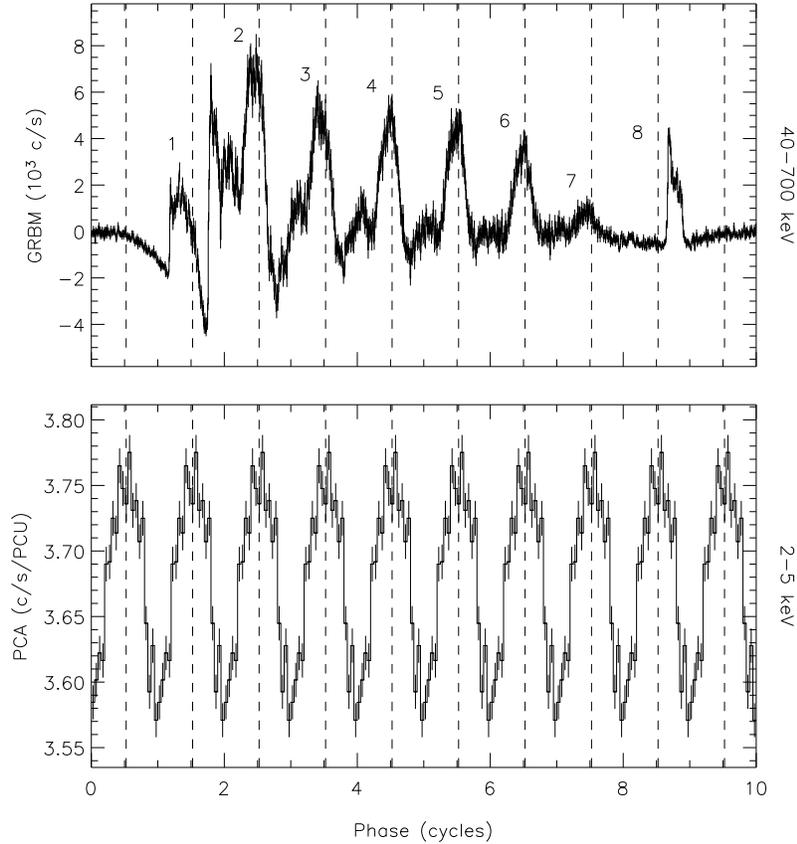,height=5.0in}}
\vspace{-0.05in}

\caption{{\it Top} -- The detrended GRBM lightcurve of the April 18 flare from
SGR~1900$+$14 (40$-$700 keV) transformed from time to rotational phase of the
star.  {\it Bottom} -- The 2$-$5 keV pulse profile from PCA data acquired
during the two weeks following this flare folded on the same ephemeris and
repeated here ten times.  Note the good agreement in the phase of the peak
between X-rays and gamma-rays.  The vertical dotted lines denotes the
centroid of the pulse peak in X-rays.}

\vspace{11pt}
\end{figure}

We next quantified the alignment of the gamma-ray to X-ray pulsations.  We
chose not to cross-correlate the profiles as was done for the X-ray data alone
due to the modest, yet significant differences in pulse shapes between the
X-ray and gamma-ray bands.  Instead, we chose to fit the peak of each pulse
profile to a quadratic and compare the centroids of the gamma-ray pulsations to
the X-ray pulse train.  We fit the peak of the X-ray profile to the quadratic
and measured its phase.  Before centroid fitting the gamma-ray pulsations, we
removed the burst envelope using a low-pass digital filter.  We identified
eight local maxima in the filtered gamma-ray lightcurve (labeled above and to
the left of each maximum in Figure 2).  We fit pulses 1 through 7 to the
quadratic function and measured their phases relative to the X-ray profile.  To
check the validity of our filtering procedure, we fit the unfiltered pulse
maxima and found that only peak 1 changed significantly.  Peak 1 is found
during the rapid and jagged rise of the flare where our filter could not
reliably subtract the burst envelope from the pulse.  We therefore omitted the
phase measurement of pulse 1.  The local maximum 8 occurs after the primary
burst envelope (encompassing pulses 1$-$7) returns to background.  We interpret
this as a separate burst event, and not a pulse.  The distinction between this
event and the other pulses is more apparent when viewing the unfiltered
lightcurve (Guidorzi et al.\ in preparation).   We find that the six remaining
gamma-ray pulsations arrive systematically earlier in phase between 0.041 and
0.093 cycles.  The average phase lag is $\sim$0.063 cycles or 23$^{\circ}$
(denoted by the triangle in Figure 1).  We conclude that there are only minor
changes in the pulse profile with both energy (2$-\sim$100 keV) and luminosity
($10^{35}-10^{42}$ ergs s$^{-1}$).  Another consequence of the relatively good
agreement between the X-ray ephemeris and the gamma-ray pulsations implies that
there was no sudden spindown following the April 18 flare of the magnitude and
type inferred for the August 27 flare.

\section{Pulse Profile and Fraction}

The X-ray pulse profile (2$-$5 keV) is quite simple showing a single, nearly
sinusoidal peak.  We investigated the evolution of the pulse shape with energy
by extracting profiles in different energy bands.  To expand the range to lower
photon energies, we also included data from {\it Chandra} Advanced CCD Imaging
Spectrometer (ACIS) observations (Kouveliotou et al.\ 2001) performed during
this epoch.  We find no significant energy dependence within the X-ray band
(0.5$-$20 keV).  Comparison of the X-ray profile with the gamma-ray profile
seen during the burst shows that he pulse shapes are similar, but not
identical.  The primary gamma-ray peak is narrower than the X-ray pulse peak
and a secondary maximum is seen at gamma-ray energies between peaks.

It was shown previously (Woods et al.\ 2001) that the dramatic change in pulse
profile found within the gamma-ray tail of the August 27 flare manifested
itself within the persistent X-ray pulse profile as well.  We searched for
changes in the X-ray pulse profile by comparing with observations from the PCA
in the year 2000.  The folded pulse profiles for the year 2000 and the April
2001 activation are shown in Figure 3.  Comparing the two pulse shapes using a
$\chi^2$ test, we find that the profiles are different at a confidence level of
99.7\%.  Specifically, the rise to pulse maximum is faster in the year 2000
profile and the decline is faster in the April 2001 profile.  We note that the
phase discrepancy of April 14 ($\gtrsim$0.24 cycles) cannot be accounted for by
the difference in shape between these pulse profiles.  A more detailed
investigation of the energy and temporal dependence of SGR pulse profiles is
presented in {G\"o\u{g}\"u\c{s}} et al.\ (2002).

\begin{figure}[!htb]
\centerline{
\psfig{file=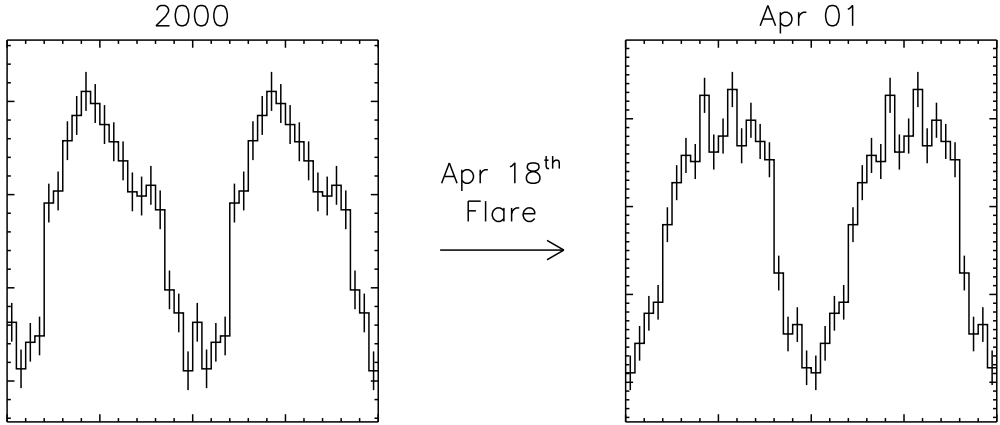,height=2.5in}}
\vspace{-0.05in}

\caption{The folded pulse profile of SGR~1900$+$14 in 2$-$10 keV X-rays as seen
with the PCA during the year 2000 ({\it left}) and during the two weeks
following the April 18 flare ({\it right}).}

\vspace{11pt}
\end{figure}

We next measured the pulse fraction following the April 18 flare using the data
from the two imaging telescopes, the {\it BeppoSAX} (SAX) MECS and {\it
Chandra} (CXO) ACIS detectors.  We measure 2$-$10 keV rms pulse fractions of
18.1(10)\%, 16.3(11)\%, 13.6(15)\%, and 12.9(12)\% for the respective epochs of
MJD TDB 52018.113 (SAX), 52021.324 (CXO), 52029.591 (SAX), and 52030.092
(CXO).  Note that the measured CXO pulse fractions differ slightly from those
reported in Kouveliotou et al.\ (2001), who reported the pulse fractions in the
0.5$-$7.0 keV energy range.  The pulse fraction of SGR~1900$+$14 increases
significantly following the April 18 flare and during the subsequent 11 days,
slowly declines, approaching its quiescent value.

We have also studied the pulse fraction evolution with energy.  The known {\it
temporal} evolution of the pulse fraction restricted our study to
contemporaneous observations (i.e.\ we could only search for changes within
individual observations).  We find no significant evolution of the pulse
fraction with energy (0.5$-$10 keV) within the individual {\it BeppoSAX} and
{\it Chandra} observations, although our limits are not very strong ($\lesssim$
25\% at 95\% confidence).

\section{Discussion}

We have observed a good phase alignment, and similarity in pulse morphology, 
between the bright gamma-ray emission ($>$15 keV) during the  April 18 flare, 
and the much fainter X-ray emission (0.5$-$10 keV)  following the flare.  These
observations have interesting implications for the source of the dissipation,
and the mechanism that shapes the pulse profile, in these two very different
states of SGR 1900+14 --  which differ by more than a factor of 10$^{6}$ in
luminosity.  

The pulse maximum in these two states is apparently produced at the same
geometric location on (or above) the surface of the neutron star. In the
magnetar model, this means either that the bursting and persistent emission are
emanating from the same region close to the star; or that the angular pattern
of the X-ray/gamma-ray emission is being modified in a similar way by resonant
(electron) cyclotron scattering high in the magnetosphere. By themselves, the
observations of the 2001 April flare and post-flare pulsations cannot
distinguish between these two models.  However, the pulse profile of SGR
1900+14 showed significant changes during the 1998 burst activation.   
Combining these earlier  observations with the measurements of the April 2001
activity, we can  begin to constrain the underyling physics.

There were gross changes in both the gamma-ray pulse morphology at high
luminosity during the 1998 August 27 flare, and in the much fainter X-ray
pulsations observed before and after this flare.  On that basis,  we argued
that the magnetic field of SGR~1900$+$14 was reconfigured  during the August 27
flare (Woods et al.\ 2001).  During the tail of that flare, the luminosity
exceeded the Eddington luminosity by as much as a factor of 10,000, which
allows a substantial optical depth to accumulate in material blown off the
surface of the star (Thompson and Duncan 1995).   It was argued that the
outburst was driven by the relaxation of a non-axisymmetric magnetic field
close to the  star, which channeled the outflow of material and created
anisotropies in the opacity high in the magnetosphere  (Feroci et al. 2001;
Thompson \& Duncan 2001). The change in pulse profile during the flare was
proposed to be  due to the contraction and simplification in topology of a hot
fireball powering the gamma-ray emission during the tail.  

A corresponding simplification was observed in the persistent X-ray pulse
profile -- from a multi-peaked shape pre-flare to a nearly  sinusoidal profile
post-flare (Woods et al.\ 2001).  In this much  lower luminosity state,  it was
hypothesized that the external magnetic field of the star retains a significant
non-potential component, and that the associated electrical currents power the
continuing non-thermal X-ray emission  (Thompson, Lyutikov \& Kulkarni 2002). 
This model implies a significant optical depth to resonant cyclotron
scattering, which is independent of frequency in the simplest case of a
self-similar and axisymmetric  magnetosphere.  The optical depth to scattering
has a maximum at the magnetic equator, and drops smoothly toward the magnetic
pole. Resonant scattering of X-rays (2$-$10 keV) by electrons  would take place
far enough from the surface of a magnetar  ($\sim5-10$ R$_{\star}$) that the
poloidal magnetic field would be almost dipolar -- hence the simplified X-ray
pulse shape  observed in all observations of the persistent emission since the
August 27 flare.  In the simplest case, this ``scattering screen'' was
activated at the onset of the giant flare when most of the energy was deposited
into the magnetosphere.  Because the quiescent X-ray spectrum was non-thermal
even before the flare, it was suggested that the outer magnetosphere of SGR
1900+14 was already twisted, and made a transition from a non-axisymmetric to
an axisymmetric configuration during the flare (Thompson et al. 2002). It
should be emphasized that, in this model, the mechanism for supplying the
scattering screen is different in the high and low  luminosity states. 
Moreover, the field geometry close to the star can remain quite complicated
even after the flare.

We now consider the observations of the 2001 April activation of SGR~1900$+$14
within the framework of the model outlined above for the August 27 flare.   The
presence of a scattering screen in the outer magnetosphere  provides a possible
explanation for the phase alignment of the pulse maxima between outburst and
quiescence.  (The optical depth in the screen is smallest along the magnetic
pole, which can be assumed to  have a fixed orientation with respect to the
body of the star.) However, the pulse profiles in the April 18 and August 27
flares are grossly different in the same energy band, in spite of having
similar  luminosities. During an outburst, the optical depth in the screen can
be temporarily augmented by ionized matter blown off the surface of star and
channeled along the magnetic field.\footnote{Only one part in  $10^{10}$ of the
energy of the August 27 flare need be expended to lift a scattering cloud that
is optically thick at 10 keV at the electron cyclotron resonance.  Similarly,
one part in a million is enough to create a large optical depth to Thomson
scattering at a  similar radius.}  As a result, one must  postulate that the
flux of material reaching a distance of 50$-$100 km  was much more irregular
during the August 27 flare than it was during the April 18 flare (whose energy
was an order of magnitude lower [Guidorzi et al.\ 2001]).  Matter and radiation
interact in a complicated way in the part of the magnetosphere where electrons
and X-rays are resonantly coupled (Thompson et al. 2002), and more detailed
work is  required to determine if this supposition is reasonable.  

An unattractive feature of this model is that the simplication in the pulse
profile observed {\it both} during the August 27 flare,  and in the persistent
emission following the flare, is ascribed to two separate physical mechanisms. 
The electrical current flowing along extended field lines is hypothesized to
have changed at the beginning of the flare, giving rise to the change in the
persistent emission pulse shape.  The pulse morphology evolution during the
flare itself is ascribed to the passive cooling of a confined fireball, with
the distribution of suspended matter undergoing a gradual simplification in
response to the changing pattern of radiative intensity emerging from the
contracting fireball.   In addition, the persistent X-ray pulse profile
observed before the flare had a similar morphology to the complicated
four-peaked pulse shape observed during the intermediate stages of the August
27 flare (Woods et al.\ 2001) -- and yet the  mechanism for supplying the
anisotropic scattering screen is different in the two cases.  One requires a
correlation between the current distribution flowing along extended field lines
in the pre-flare state, and the energy release within the flare itself.

An extreme alternative to this model is that the {\it entire} external magnetic
field of SGR~1900$+$14 was reconfigured into a simpler geometry over the
duration of the August 27 flare (Woods et al.\ 2001).  In this model, the
magnetosphere is assumed to remain optically thin post-flare (i.e.\ no
scattering screen). The similar morphology and good phase alignment of the
burst and post-burst pulsations during the 2001 April 18 flare follow naturally
if there is a common dissipative region.  However, the reconfiguration of the
field would have to progress on a timescale  $\sim$10$^7$ times longer than the
Alfv\'en crossing time of the magnetosphere ($\sim$30 $\mu$sec), and $\sim
10^4$ times longer than the corresponding timescale in the deep interior of the
star (e.g. Thompson \& Duncan 1995).   Recently, Lyutikov (2003) argued that
reconnection in the magnetosphere driven by a tearing mode instability would
proceed on a longer timescale of order $\sim$10 ms. This timescale is
consistent with the rise time of the more common SGR bursts
({G\"o\u{g}\"u\c{s}} et al.\ 2001), but still many orders of magnitude shorter
than the timescale of the pulse profile change observed during the August 27
flare ($\sim$5 min).  

A global magnetospheric reconfiguration during the August 27 flare has the
advantage of providing the most straightforward explanation of the pulse
morphology changes seen in SGR~1900$+$14, both in 1998 and 2001.  However,
there is still the challenge of explaining how the field would continue to
reconfigure itself over a duration of $\sim 400$ seconds in a very smooth
manner -- without inducing secondary instabilities and reconnection events that
would manifest themselves as sudden changes in the X-ray flux.  A  compromise
between these models would involve a gradual simplication of the most extended
magnetic field lines (those which extend out to the electron cyclotron
resonance for soft gamma-rays up to $\sim$100 keV).

We have shown that the torque in the aftermath of the 2001 April 18 flare was a
factor $\sim$3 larger than the torque measured immediately following the 1998
August 27 flare.  In contrast, the burst activity following the August 27 flare
was more intense and persisted longer than the activity following April 18. 
The torque measured following this intermediate flare is large as compared to
the post-August 27 torque, but is not the largest measured for SGR~1900$+$14. 
Comparison of the burst activity and spindown in April 2001 with previous
epochs in SGR~1900$+$14 strengthens our earlier conclusion (Woods et al.\
1999b, 2002) that there exists no {\it direct} correlation between burst
activity and torque in this system.

Our comparison of the gamma-ray pulse profile with the X-ray ephemeris has
shown that there is no evidence for transient post-flare spindown of the
magnitude and type inferred for the August 27 flare.  This observation also
lends credence to the interpretation of the phase mis-alignment in the August
27 flare as being due to a sudden spindown event (Palmer 2001; Woods et al.\
1999b) rather than a strong energy dependence in the pulse profile.  Although
we can exclude a sudden spindown like that inferred for the August 27 flare,
the fortuitous observation preceding the April 18 flare indicates that there
was some timing anomaly near the epoch of the flare.  Unfortunately, the {\it
exact} manner in which the spin frequency of the star evolved leading up to and
through both flares is not certain due to uncertainty in the pre-flare spin
ephemerides.  The value of precise pre-flare spin ephemerides is best
illustrated in the recent SGR-like outburst from the anomalous X-ray pulsar
1E~2259$+$586 (Kaspi et al.\ 2003).  In this case, the continuous monitoring of
1E~2259$+$586 leading up to this outburst allowed Kaspi et al.\ (2003) to
detect a sudden spin-up or glitch that coincided with the burst activation. 
Since the onset of burst activity in SGRs is unpredictable, continuous
monitoring of their spin ephemerides is required to quantify timing events such
as the glitch detected in 1E~2259$+$586 coincident with bursts, and hence
constrain the underlying physical mechanism responsible for the busts.

\acknowledgments{\noindent {\it Acknowledgements} -- We thank the RXTE/SDC and
the SAX/SDC for pre-processing the RXTE/PCA and BeppoSAX data, respectively. 
This work was funded through a NASA ADP grant (NAG 5-11608) for PMW and a Long
Term Space Astrophysics program (NAG 5-9350) for both PMW and CK.  MHF and EG
acknowledge support from the cooperative agreement NCC 8-200.}

\end{document}